# Grayscale Electron Beam Lithography Direct Patterned Antimony Sulfide

Wei Wang[1,2*], Uwe Hübner[1], Tao Chen[1,2], Anne Gärtner[3], Joseph Köbel[1,2], Franka Jahn[1], Henrik Schneidwind[1], Andrea Dellith[1], Jan Dellith[1], Torsten Wieduwilt[1], Matthias Zeisberger[1], Tanveer Ahmed Shaik[1,2], Astrid Bingel[3], Markus A Schmidt[1,4,5], Jer-Shing Huang[1,2,6,7], Volker Deckert[1,2,4,8*]

w.wang@uni-jena.de,

volker.deckert@uni-jena.de

**Abstract**

The rise of micro/nanooptics and lab-on-chip devices demands the fabrication of three-dimensional structures with decent resolution. Here, we demonstrate the combination of grayscale electron beam lithography and direct forming methodology to fabricate antimony sulfide structures with free form for the first time. The refractive index of the electron beam patterned structure was calculated based on an optimization algorithm that is combined with genetic algorithm and transfer matrix method. By adopting electron irradiation with variable doses, 4-level Fresnel Zone Plates and metalens were produced and characterized. This method can be used for the fabrication of three-dimensional diffractive optical elements and metasurfaces in a single step manner.

**Introduction**

Based on refraction and reflection, the conventional optical elements (OEs), such as lenses, mirrors, prisms, waveplates, etc., have been widely adopted in industrial application and scientific research for hundreds of years. Nowadays, those OEs are well engineered and can bring performance close to the theoretical prediction, such as Abbe diffraction limit [1, 2]. However, conventional OEs are usually heavy, bulky and often require complex manufacturing procedure. These disadvantages make them not ideal for the emerging fields of integrated nano/biophotonics systems, such as lab-on-chip devices and point of care diagnostics. As these fields would prefer light weight, ultra-thin, higher performance and low cost OEs. Diffractive optical elements (DOEs) and optical metasurfaces are two promising candidates to fulfill these requirements. DOE defines a microstructure surface which is used to modulate the wave front of incident light via diffraction, instead of refraction. Classic DOEs, such as Blazed grating [3], Spiral Phase Plates (SPPs) [4], and Fresnel Zone Plates (FZPs) [5] have already widely used in consumer products and scientific instruments. Optical metasurface is an artificial nanostructured interface consisting of designed building blocks arranged on a subwavelength scale. The building blocks are usually consisting of plasmonic or dielectric materials, which can directly modify light properties such as amplitude, phase, and polarization.

[1] Leibniz Institute of Photonic Technology (IPHT), 07745 Jena, Germany.
[2] Institute of Physical Chemistry, Friedrich Schiller University Jena (FSU), 07743 Jena, Germany
[3] Functional Optical Surfaces and Coatings, Fraunhofer Institute for Applied Optics and Precision Engineering (Fraunhofer IOF), 07745 Jena, Germany
[4] Abbe Center of Photonics and Faculty of Physics, FSU, 07745 Jena, Germany
[5] Otto Schott Institute of Material Research, FSU, 07745 Jena, Germany
[6] Research Center for Applied Sciences, Academia Sinica, 11529 Taipei, Taiwan
[7] Department of Electrophysics, National Yang Ming Chiao Tung University, 30010 Hsinchu, Taiwan
[8] Jena Center for Soft Matter (JCSM), FSU, 07743 Jena, Germany

In the past few years, metasurfaces have attracted great attention due to the superior functionalities and performances that are hard to achieve by conventional OEs [6-12].

The vast application of DOEs and metasurfaces drives the ever-increasing demand for high accuracy micro- and nanostructure manufacturing. High-precision mechanical milling, such as single point diamond turning, can be adopted for the production of free form components for DOEs [13]. However, limited resolution hinders the scalability of such mechanical fabrication technology. Electron beam lithography (EBL) is a very accurate technique for creating ultra-fine patterns as it is based on the principle of the scanning electron microscope (SEM) and thus inherits the high-resolution merit of SEM. Currently, this method is the mostly adopted approach for DOEs and metasurfaces fabrication. The EBL functioning principle is very simple: a focused or shaped beam of electrons is scanned across a surface covered (normally through spin-coating) by an electron-sensitive material (electron resist). Upon electron beam exposure, the positive resists [14] (polyolefinsulfones, polymethylmethacrylate (PMMA), polyisocyanates, etc.) degrade and negative resists [15] (epoxy-based SU-8, poly (methyl vinyl ketone, poly(vinyl methyl ether), poly(vinyl alcohol), etc.) polymerize according to designed patterns. The unexposed positive resists have a very low solubility, while the exposed areas are soluble in developer solution. Negative resists behave in the opposite manner. After the development process, the pattern composed of resists can be used to transfer into designed materials via etching. DOEs and metasurfaces are normally fabricated through the deposition or etching of the target substance, such as noble metals [16, 17] or dielectric materials [9, 18]. Grayscale electron beam lithography (g-EBL) is a fabrication technology that combines both nanometer scale resolution and free form topography. For some electron resists, beside high lateral resolution, tunable thickness can also be obtained via different exposure dose, i.e., grayscale lithography (Fig.1a).

Recently, antimony trisulfide ($Sb_2S_3$) has attracted much attention as it is a promising candidate for the fabrication of lab-on-chip devices in the IR and mid-IR regions [19-22], due to its high refractive index and low loss, inherent infrared transparency, high nonlinearity, semiconducting nature and topological effect. Compared with other commonly used optical material (such as PMMA, SU-8, BK7, etc.), $Sb_2S_3$ has much higher refractive index (n~2.5 and 4.5 at 633 nm vacuum wavelength for amorphous and crystalline $Sb_2S_3$) while still maintaining low losses (Fig. 2d). Such high refractive index and low loss nature makes $Sb_2S_3$ an ideal candidate for the fabrication of high performance DOEs and metasurfaces. What is more, $Sb_2S_3$ is also a Phase Change Material (PCM), which has at least two solid phases (amorphous and crystalline) with substantially different optical properties, especially the refractive index. Both of these phases remain stable at room temperature, and can be transformed from one to the other when subjected to a thermal impulse, which can be supplied through electrical [22] or optical means [21, 23]. This transformation can be accomplished within nanosecond timescales, rendering $Sb_2S_3$ highly appealing for the development of dynamic DOEs and metasurfaces [24], offering precise control over output properties. Moreover, the material allows for the creation of intermediate states, situated between full amorphous and fully crystalline phases, thereby adding further versatility and functionality to its applications. In this work, we present a novel method for the genuine 3D patterning of high refractive $Sb_2S_3$ with g-EBL.

## Part I: g-EBL Patterning mechanism of $Sb_2S_3$

The basis of this work is an antimony-butyldithiocarbamic acid (Sb-BDCA) based molecular precursor solution. The Sb-BDCA complex was synthesized by the reaction of carbon disulfide ($CS_2$), n-butylamine and antimony (III) oxide ($Sb_2O_3$). According to a previous study [25] and nuclear magnetic resonance (NMR) analysis (Fig. S1 in Supplement Info part SI), the reaction of $CS_2$ and n-butylamine will form butylcarbamodithioacid (see Scheme. 1a)

The BDCA can be used to dissolve the metal oxides ($M_xO_y$) and metal hydroxides ($M(OH)_x$) to form metal sulfide ($M_xS_y$) precursor solutions. The thionothionic acid group in BDCA will react with the $M_xO_y$ and $M(OH)_x$ to form a metal-organic complex, known as $M(BDCA)_x$ (see Scheme. 1b). Upon external stimuli, such as heating or electron beam irradiation, the metal-BDCA will decomposed into associated metal chalcogenide. Such metal-carbamodithioacid precursor-based metal chalcogenide preparation method was first developed by Nomra, etal. in 1988 [25] as an effort to obtain indium sulfide for optoelectronic applications. In recent years, Gang Wang etal. utilized the same concept to obtain copper indium gallium selenide (CIGSe) thin film for the fabrication of higher efficiency solar cell [26] ($\eta$=8.75 % at AM 1.5G). In 2017, Xiaomin Wang etal. reported the fabrication of $Sb_2S_3$ thin film via the thermal induced decomposition of Sb-BDCA for the first time [27]. Such solution-based metal chalcogenide production route has great scientific and industrial application potential. As BDCA is inexpensive, relatively low toxic and easy to make.

In this previous study [28], Wei Wang etal. demonstrated for the first time that the metal-BDCA complex can also be used for the direct patterning of metal chalcogenide nanostructures via various lithography methods, including EBL, ultraviolet photolithography, thermal scanning probe lithography and two-photon absorption lithography. Take the direct patterning of $Sb_2S_3$ with EBL for example. The Sb-BDCA complex in ethanol solvent was first spin-coated on the conductive substrate to form a homogeneous thin film. Upon electron beam irradiation, the Sb-BDCA complex was decomposed, thus forming amorphous $Sb_2S_3$ in the designed patterns as a result (Fig. 1a in [28]). The unexposed area is still composed of Sb-BDCA complex and can be washed off with ethanol or isopropanol during the development process. After development, only $Sb_2S_3$ in the designed patterns were left on the substrate. Thus, the Sb-BDCA complex resembles SU-8 and can be considered as negative type resist. $Sb_2S_3$ nanostructures with 50 nm lateral resolution and high fidelity with design pattern were fabricated with EBL (Fig. 2 in [28]).

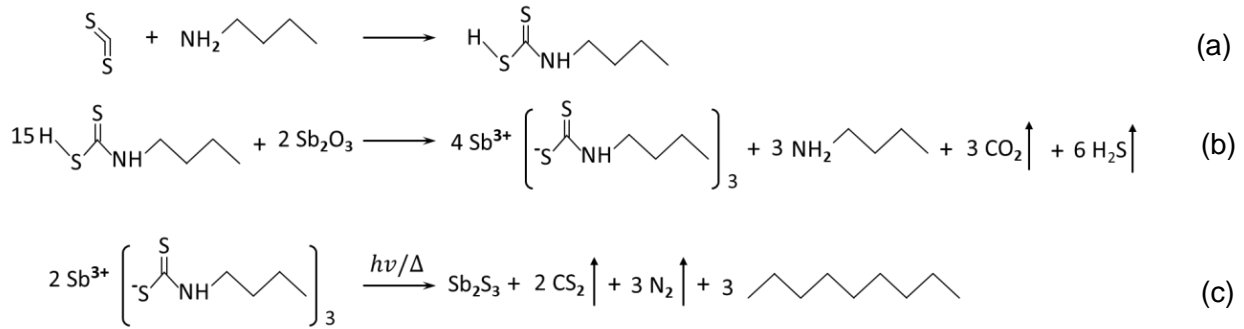

**Scheme 1**: (a) Formation of BDCA via the reaction of $CS_2$ and n-butylamine. (b) Formation of Sb-BDCA complex via the reaction of $Sb_2O_3$ and BDCA [29] [30]. (c) Formation $Sb_2S_3$ of via the decomposition of Sb-BDCA complex.

In the recent study, the authors found that the thickness of the EBL patterned $Sb_2S_3$ structure can be tuned by electron beam irradiation dose. That is to say, $Sb_2S_3$ nanostructures with controlled lateral and vertical profile can be both obtained through a single step g-EBL direct patterning scheme. The procedure chain for the production of grayscale $Sb_2S_3$ structure is illustrated in Fig. 1a. The Sb-BDCA complex was first spin-coated on the target substrate (Fig. 1a (i)). As it illustrates in Fig. 1a(ii), the grayscale lithography is realized by varying the electron beam irradiation dose ranging from 2000 to 30000 µC/cm$^2$ on different parts of the Sb-BDCA film. Sb$_2$S$_3$ patterns with desired height profile will be obtained after development. As it shown in Fig. 1a(iii), the resulting thickness of the pattern is proportional to the irradiation dose, i.e., the thickness increases with higher dose. Fig. 1b shows the scanning electron microscopy (SEM) image of an 8-level staircase structure array. A dose variation from 8000 to 15000 µC/cm$^2$ with 1000 µC/cm$^2$ step was performed in order to determine the profile of the staircase structure. The exposure steps can be clearly seen in the right inserted figure. The bottom inserted figure of Fig. 1b depicts the cross section of the stair structure array, which proves the high fidelity as well reproducibility of the Sb-BDCA based g-EBL method. Blazed grating, one of the widely adopted DOEs, is actually composed of such staircase array with designed periodicity and blaze angle [31]. In Fig. 1c, the atomic force microscopy (AFM) micrograph of a spiral phase plate (SPP) like structure fabricated with g-EBL is displayed. SPP is also a commonly used DOEs, which is composed of transparent elements with thickness increases around the central axis [4]. The Sb$_2$S$_3$ composed SPP like structure exhibits a very low roughness, with RRMS<3 nm (average RMS on a surface area of 2 x 2 µm$^2$ area) as well as sharp spiral phase step and high definition of the central singularity, indicating a high quality SPP like structure. A detailed discussion regarding the performance of the Sb-BDCA in negative resist perspective and comparison of the key parameters with typical commercial electron resists can be found in Supplement Info Part SII.

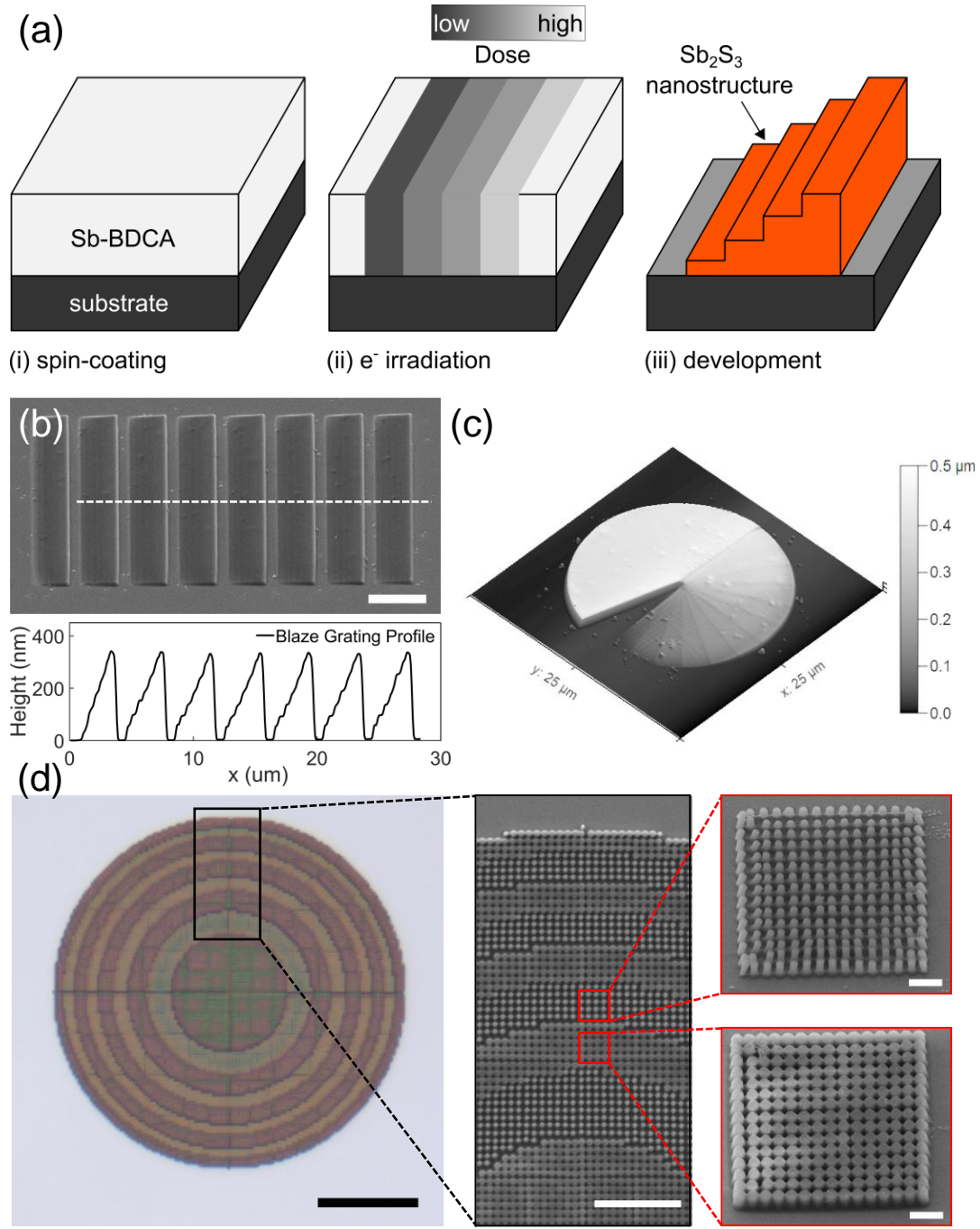

**Figure 1.** (a) Scheme of the grayscale EBL process steps. From left to right: (i) Sb-BDCA complex thin film deposited on the substrate via spin-coating. (ii) Electron beam irradiation of the Sb-BDCA complex using different doses. (iii) Reveal the grayscale $Sb_2S_3$ structure after development. (b) SEM micrograph of 8-level stair structures array. Scale bar, 5 μm. The bottom insert figure shows surface cross section along the white dashed line of the stair structure. (c) AFM micrograph of 24-levels spiral phase plate like structure. (d) Left: Optical microscopy image of a fabricated propagation phase metalens based on cylindrical $Sb_2S_3$ pillars. Middle: SEM image of a 2-level $Sb_2S_3$ metasurface. Right: Nanopillar $Sb_2S_3$ meta-atoms arrays with 450 nm periodicity as building blocks of the metasurface. Scale bar: from left to right, 15 μm, 5 μm and 1 μm.

**Part II: Refractive index determination**

The refractive index n=n+ik, is the most important parameter for the design and fabrication of DOEs and metasurfaces [32]. The application of metasurfaces, for instance, was initially based on metallic subwavelength structures [33]. However, due to the large absorption (k, the imaginary part of the refractive index) in metals, it was not possible to create high efficiency meta devices in the visible and near-infrared (NIR) domain. This promoted the fabrication of metasurfaces using dielectric materials due to low optical losses. For devices with a high numerical aperture (NA>0.6), it is ideal to use dielectric materials with a high refractive index (n>2), such as Si, $TiO_2$, GaN and SiN, as these materials will bring higher efficiency than low refractive index materials (n<2), such as PMMA, SU-8 [32] [34], and $SiO_2$ [35].

Thus, the precise determination of the complex refractive index of the EBL pattern $Sb_2S_3$ thin film is vital for the design and fabrication of associated DOEs and metasurfaces [32]. The spectroscopic ellipsometry method is the most commonly applied method for determining the refractive index and thickness of optical materials [36]. However, conventional macro ellipsometry requires a minimum sample area of 2×2 $mm^2$ for effective data acquisition. Due to the very low sensitivity of the Sb-BDCA resist (see Fig. S2b and Table S1), it would consume at least 516 hours to fabricate such big size on Si substrate with 100 pA electron beam current. In this work, we investigated the refractive index of EBL patterned $Sb_2S_3$ through the reflectance spectra. As it schematically depicts in Fig. 2a, EBL direct patterned 100×100 µ$m^2$ sized $Sb_2S_3$ squares with varied thickness (d) were fabricated on Si substrate. The normal incident reflectivity spectra (in the range of 400 nm to 1000 nm) of the $Sb_2S_3$ squares were recorded with an optical microscopy equipped with a microspectrophotometer, which enables the measurement of reflectivity of extremely small areas that are not possible for ordinary spectrophotometers. Fig. 2c depicts the reflectance spectra (solid line) of same batch $Sb_2S_3$ squares without and with 180 °C thermal treatment. The reflectance varies periodically with wavelength due to the constructive and destructive interference of the waves reflected from the front and rear surfaces of the $Sb_2S_3$ film. The color dependence of the thickness of the $Sb_2S_3$ squares were also recorded and presented in the right side image of Fig. 2c. On one hand, the positions of the minima and maxima positions in the reflectance spectra clearly show strong dependence of the thickness. More specifically, the minima positions gradually shift to longer wavelength (red shift) with increased sample thickness in an ordinary manner, which is a typical behavior of dielectric materials [37] and agrees with the previous report [23]. On the other hand, the reflectance spectra and color difference between unannealed and annealed sample is substantial. This phenomenon can be attributed to the different thicknesses, as the thicknesses of the $Sb_2S_3$ squares were reduced by almost 50% after annealing at 180 °C. The spectra and color difference can also be caused by the optical property change before and after thermal treatment.

In this work, we developed a method utilizing a computational genetic algorithm (GA) and transfer matrix method (TMM) to calculate the complex refractive index of $Sb_2S_3$ based on the experimentally measured reflectivity spectra of the thin films. A more detailed discussion of this newly developed refractive index determination method is given in Supplement Info Part SIII and all associated information will be presented in a separate publication. Fig. 2b depicts the flowchart of this GA and TMM based refractive index optimization algorithm. The interference effect in a multilayer planar stack (air-$Sb_2S_3$-Si, as shown in Fig. 2a) is modeled with TMM, which can provide reliable information regarding the distribution of the electric magnetic field and more importantly, the reflection and transmission spectra [38, 39]. The electric magnetic field was incident from the air side toward the Si. Air and Si layers were treated as semi-infinite transparent ambient and semi-infinite substrates. The free parameters of the TMM model are the refractive index (n(λ)) and the extinction factor (k(λ)), as the thicknesses of the $Sb_2S_3$ were already obtained by experiment (Fig. 2c). The experimentally obtained reflectance spectra were used as fitting target. TMM simulation, with varying n(λ) and k(λ) values, was performed to find the refractive index values that best fit the experiment data (Fig. 2c). GA is very well suited to solve such a parameter optimization problem since it performs a random search in the parameter space with a weak dependence on initial values and its ability to find global minima with few iterations [40]. As the number of the free parameters is two (n(λ) and k(λ)), thus at least two reflectance spectra with different thicknesses should be used as fitting targets to obtain convergent results (Fig. 2b). As it depicts in Fig.2c, a good agreement was found between the experimentally obtained reflectance spectra (solid line) and the TMM simulation (dashed line) with the optimized n(λ)/k(λ) values given by the as mentioned optimization algorithm. Thus, the $Sb_2S_3$ films obtained with the same heat treatment (i.e., unannealed, 180 °C annealed or 300 °C annealed), but with different thicknesses all have the same refractive index, indicating that the refractive index of $Sb_2S_3$ is independent of thickness.

The complex refractive index of $Sb_2S_3$, patterned and annealed through EBL, was accurately determined by analyzing reflection spectra in this study. The parameter optimization method employed, a combination of GA and TMM, is a numerical approach known for discretizing problems and utilizing algorithms to derive approximate solutions. However, when extending our inquiry beyond the defined spectrum domain (400 nm to 1000 nm wavelength), an analytical approach becomes indispensable. Analytical analysis offers precise solutions, contributing to a deeper understanding of the mathematical relationships and physical mechanisms governing the system. Furthermore, the process of attaining an analytical solution often provides valuable insights into the system's behavior, facilitating interpretation and generalization. The Tauc-Lorentz (TL) oscillator [41, 42] stands out as the predominant model for metal chalcogenides, widely employed for its efficacy in describing the dispersion relationship of dielectric and semiconductor materials. Originating from the work of Jellison and Modine [41], this model is particularly adept at characterizing materials that exclusively absorb light above their optical bandgap. Its successful application spans various substances, including amorphous Si [43] and metal chalcogenides like $As_2S_3$ [44], $Sb_2S_3$ [21, 22, 45], and $Sb_2Se_3$ [21, 22]. For a comprehensive understanding of the TL model, readers can refer to the detailed description provided in Supplement Info Part SIV.

The complex refractive index of the EBL patterned $Sb_2S_3$ based on TL model were given in Fig. 2d and Fig. S7. TL parameters of the EBL patterned $Sb_2S_3$ were given in Table. S2. As shown in Fig. 2d, both the refractive index and extinction coefficiency of $Sb_2S_3$, increased with annealing temperature. Such positive correlation between annealing temperature and complex refractive index of $Sb_2S_3$ has been shown by Gutierrez et al. [46] and Delaney et al. [21]. The refractive index of the 180 °C annealed $Sb_2S_3$ reported in this work is close to the value of the amorphous $Sb_2S_3$ (a-$Sb_2S_3$) prepared by chemical bath deposition (CBD) [46] and electrophoretic deposition method (EPD) [47]. However, differences in the values of refractive index are also reported. For radio frequency sputtered amorphous $Sb_2S_3$ films [22], a refractive index of n=3.14 at 633 nm was determined, in contrast to the value of n =2.60 reported in here, indicating the effect of the deposition method on the density of the amorphous $Sb_2S_3$ film. The refractive index of the 300 °C annealed $Sb_2S_3$ calculated in here is close to the value of crystalline $Sb_2S_3$ (c-$Sb_2S_3$) prepared by thermal annealing of the amorphous $Sb_2S_3$ [48]. From the TL parameterization (Table. S2), the optical band gap was determined to be 2.33, 1.91 and 1.45 eV for the unannealed, 180 °C and 300 °C annealed $Sb_2S_3$ respectively. The band gap values of 180 °C and 300 °C annealed $Sb_2S_3$ are in good agreement with those report by Juneja et al. [49] and Delaney et al. [21] [ref], i.e., 1.97 eV and 1.45 eV for a- and c-$Sb_2S_3$ respectively. The 2.33 eV band gap of unannealed $Sb_2S_3$ is close to the value (2.2±0.1 eV) of a- $Sb_2S_3$ prepared by sputtering [21]. To confirm that the change of the refractive index of the EBL patterned $Sb_2S_3$ are caused by the thermal induced phase change, i.e., changing from amorphous to crystalline state, Raman spectra (Fig. 2e) were taken at various annealing temperatures. For unannealed and 180 °C annealed samples, the Raman spectra show several broad peaks, which is the characteristic feature of amorphous $Sb_2S_3$ [21, 23, 45, 50]. For amorphous $Sb_2S_3$, the features at 130 cm$^{-1}$ and 300 cm$^{-1}$ correspond to the Sb-S and S=S vibrational modes [21]. After 300 °C annealing, amorphous $Sb_2S_3$ crystallizes in an orthorhombic structure with space group symmetry Pnma (no. 62) with lattice parameters a = 1.1302 nm, b = 0.3834 nm, and c = 1.1222 nm [51, 52]. The Raman spectra of c-$Sb_2S_3$ exhibits multiple sharper peaks, which indicates that a crystalline structure is presented. For crystalline $Sb_2S_3$, the vibrational modes of symmetry groups $A_g$ at 125 cm$^{-1}$, 154 cm$^{-1}$, 284 cm$^{-1}$ and 312 cm$^{-1}$, $B_{1g}/B_{3g}$ at 240 cm$^{-1}$ and $B_{2g}$ at 188 cm$^{-1}$ [50, 53]. The $A_g$ mode at 240 cm$^{-1}$ represent the symmetry S-Sb-S bending mode. While the 284 cm$^{-1}$ and 312 cm$^{-1}$ anti-symmetric and symmetric S-Sb-S stretching modes can be attributed to the trigonal pyramidal vibration modes of $SbS_3$ units [45, 53].

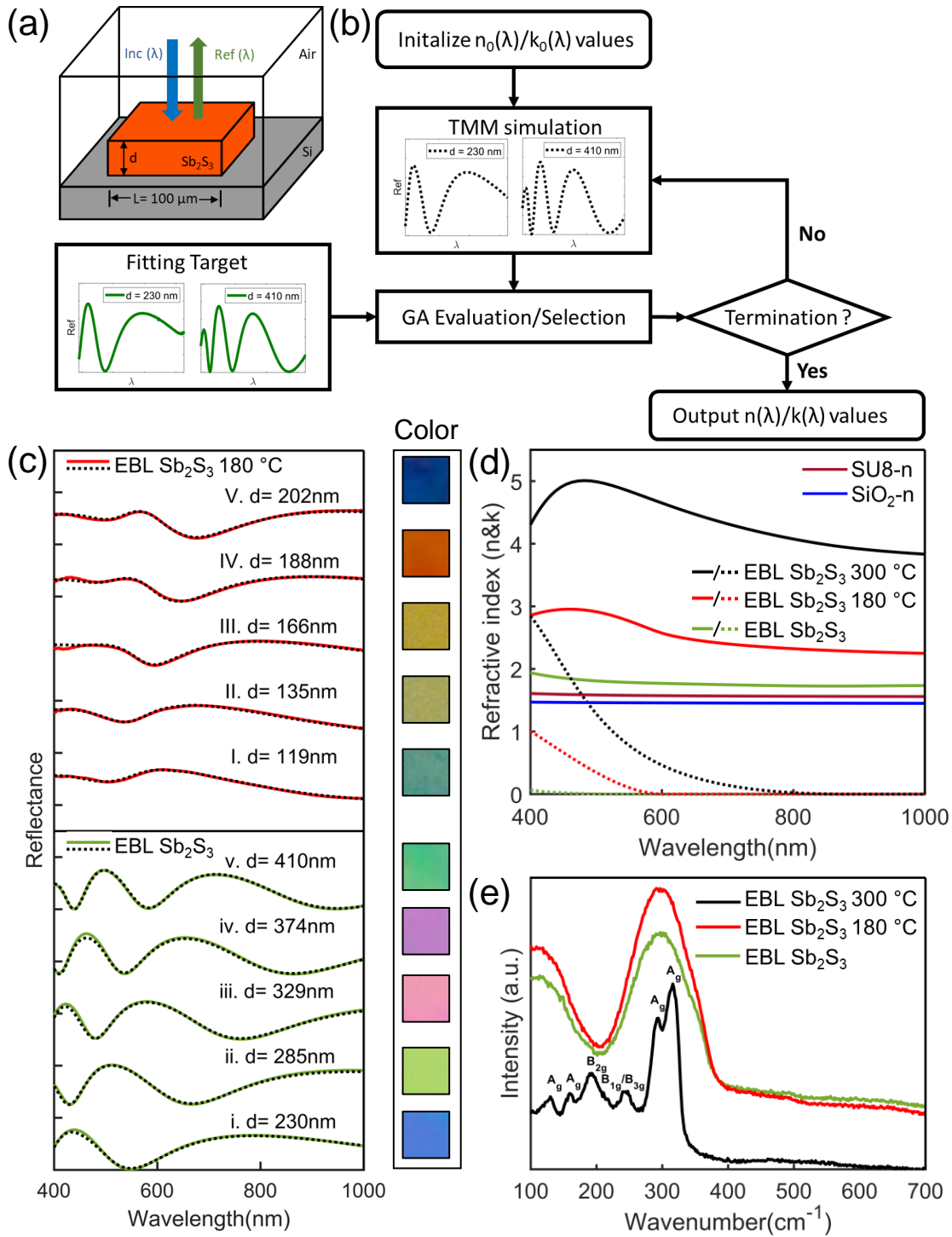

**Figure 2.** (a) Scheme of the model for microspectrophotometer measurement. (b) Flowchart of this GA and TMM based refractive index optimization algorithm. (c) Reflectance spectra of unannealed and 180 °C annealed EBL patterned $Sb_2S_3$. Solid line: experimentally obtained reflection spectrum. Dashed line: calculated reflection spectrum. (d) The comparison of the refractive index (n, solid line, k, dashed line) between unannealed, 180 °C, and 300 °C annealed EBL patterned $Sb_2S_3$, SU-8 [34], and fused-silica ($SiO_2$) [35]. (e) Raman spectroscopy for unannealed, 180 °C, and 300 °C annealed EBL patterned $Sb_2S_3$.

**Part III: 3D FZP and metalens demonstration:**

| | | $E_g$ (eV) | @ 532 nm | @ 633 nm | @ 800 nm | @ 1000 nm |
|---|---|---|---|---|---|---|
| Sb$_2$S$_3$ | EBL | 2.33 | 1.790+0i | 1.762+0i | 1.730+0i | 1.734+0i |
| | EBL 180 °C | 1.91 | 2.871+0.197i | 2.604+0i | 2.426+0i | 2.355+0i |
| | EBL 300 °C | 1.45 | 5.001+0.978i | 4.560+0.310i | 4.043+0.011i | 3.793+0i |
| SU-8 [34] | | - | 1.580+0i | 1.571+0i | 1.563+0i | 1.559+0i |
| SiO$_2$ [35] | | - | 1.460+0i | 1.457+0i | 1.453+0i | 1.450+0i |
| TiO$_2$ [54] | | 3.17 | 2.450+0i | 2.389+0i | 2.341+0i | 2.313+0i |
| Si [55] | | 1.66 | 4.137+0.034i | 3.864+0.016i | 3.669+0.005i | 3.575+0i |

**Table 1.** Values of refractive index n=n+ik and optical band gap ($E_g$) of EBL patterned Sb$_2$S$_3$, SU-8 [34], SiO$_2$ [35], TiO$_2$ [54] and Si [55], at selected vacuum wavelength of 532, 633, 800, and 1000 nm.

FZP and metalens are optical devices ideal for the miniaturization of imaging and focusing instruments, but they have distinct differences in terms of design, working principles, and applications. Both FZPs and metalenses are designed to manipulate the wavefront of light, allowing for the control of phase and amplitude. Thus, both must provide a similar phase shift profile as their equivalent refractive counterpart. The FZP is a subcategory of DOEs that functionalized by the diffraction patterns created by the arrangement of concentric rings or zones, exploiting the Fresnel diffraction phenomenon [56]. Metalens, a subcategory of metasurfaces, is a transmission [57, 58] or reflection [59, 60] device that relies on subwavelength metallic or dielectric [57] structures to achieve phase modulation, allowing for precise control over the wavefront of light. Despite difference operation mechanism, both FZP and metalens should satisfy the spatial variation of the phase as a parabolic lens [61]:

$$\varphi(x,y) = \left[-\frac{2\pi n_{medium}}{\lambda_0}\left(\sqrt{x^2+y^2+f_0^2}-f_0\right)\right]_{2\pi} \quad (1)$$

Where $\lambda_0$ is working wavelength in vacuum, $f_0$ is the designed focal length, $n_{medium}$ is the refractive index of the working media, x and y are spatial coordinates on the lens. The modulo-$2\pi$ operation converts the convert the continuous phase profile of a parabolic lens into phase-wrapped (blazed) zones whose phase variation lies between [0, $2\pi$]. The plot of the unwrapped and the $2\pi$-wrapped phase profiles are shown in Fig. S8a(ii) and Fig. 3a (dashed line), respectively. Under this ideal phase distribution, the incident plane wavefront can be perfectly transformed into a spherical wavefront.

For FZP, owing to manufacturing difficulties of the continuous surface profile, discrete approximation of the continuous phase-wrapped lens, i.e., multilevel ($g$-level, $g$ is the number of phase levels) FZPs are designed and fabricated in practice [62, 63]. The solid line in Fig. 3a depicts the 4-level approximation (also shown in Fig. S8a(iv)), which has a theoretical diffraction (focusing) efficiency up to 81.2 % according to the simulation (see Supplement Info Part SV for more detail).

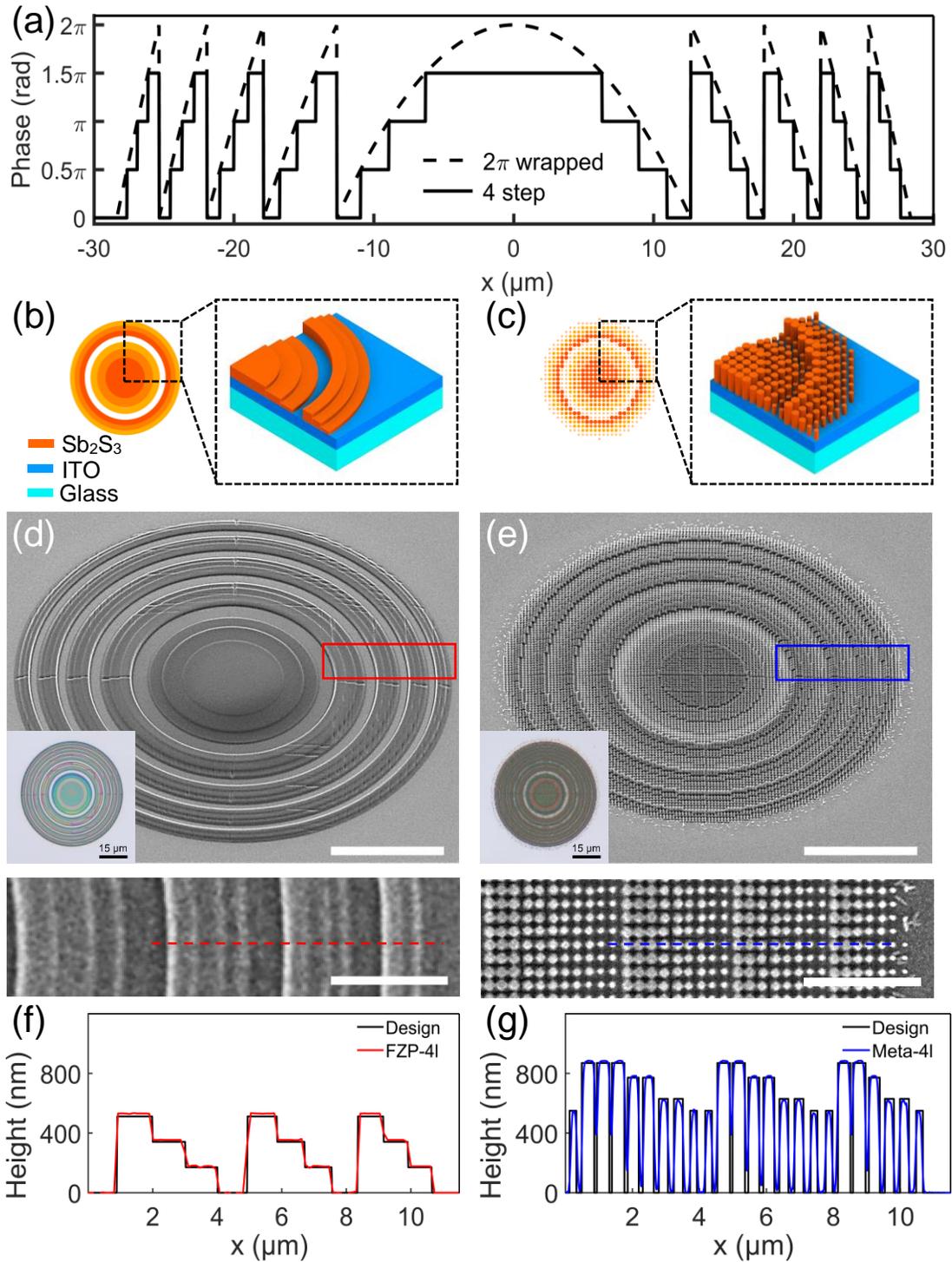

**Figure 3.** (a) 2π-wrapped phase profile of a planoconvex (positive or focusing) lens with $f_0$ =150 μm and NA = 0.19 at 532 nm vacuum wavelength (dashed line) and its 4-level approximation (solid line). Schematic of the constituent element of 4-level $Sb_2S_3$ (b) FZP and (c) metalens. SEM and optical microscope images of 4-level $Sb_2S_3$ FZP (d) and metalens (e) Scale bar: 15 μm, 4 μm, top and down. Surface cross section of $Sb_2S_3$ FZP (f) and metalens (g) along the dashed line shown in (d) and (e).

For metalens that focuses a normally incident plane wave to a diffraction-limited spot on the focal plane, the ideal phase distribution also follows Eq. (1). Due to the finite periodicity, it is fundamentally not possible to achieve the ideal phase profile with metalens [64]. This means whatever the target phase might be, the metalens design involves translating the continuous spatial phase profile defined in Eq. (1) into a discrete distribution. The deviation from ideal phase profile making the transformed wavefront of metalens be deformed from the ideal spherical wavefront. The wavefront aberration increases more and more toward the edge of high numerical aperature (NA) metalens, as the phase gradient is too high to be realized by metaatom with finite physical size [64]. It's imperative to maintain a nanostructure array period (metaatom spacing) smaller than the Nyquist sampling limit ($\lambda_0/(2NA)$) to mitigate higher order diffraction loss. For metasurface, the effective index of the building block material is decreased ($n_{eff} < n_{material}$) due to the smaller cross-section ($\sim \frac{\pi \lambda^2}{4}$), leading to an increased thickness to cover the relative phase change of $2\pi$. Such ultra-thick thickness makes conventional polymer based electron resists not suitable for the exciting and emerging fields of high performance OEs. A detailed discussion regarding the multilevel FZP design and characterization can be found in Supplement Info Part SIV.

Table 1 summarizes refractive index of EBL patterned $Sb_2S_3$ and some common optical materials for DOEs and metasurfaces application. The optical indices of SU-8 and $SiO_2$ were also given in Fig. 2d as comparison. SU-8 and PMMA are two of the most commonly used resists for such the fabrication of multilevel DOEs [4]. These resists have been successfully adopted for the fabrication of blazed gratings [3], SPPs [4], and FZPs [5] with g-EBL direct patterning. SU-8 is mostly adopted for gray scale DOEs fabrication [65-67], while Si and $TiO_2$ are most commonly used for metasurfaces. However, the relatively low refractive index of SU-8 and PMMA ($n_d$=1.492 and 1.590 for PMMA and SU-8 respectively [68, 69]) is considered as a major disadvantage. Take the fabrication of $g$-level FZP for example, as an effort to achieve designed phase modulation at 532 nm, the thickness range $d$ can be calculated with the following equation [61]:

$$d = \frac{\lambda_0}{(n_{material} - n_{medium})} \times \frac{g-1}{g} \qquad (2)$$

where $n_{material}$ is the refractive index of the optical material. Thus, for PMMA, the required thickness range should be 0.811 µm if the working media is air ($n_{air}$=1), and 2.46 µm in water ($n_{water}$=1.33) for $g = 4$. Compared with SU-8 and PMMA, the EBL patterned $Sb_2S_3$ exhibits higher refractive index and still maintain no loss in VIS-IR spectrum. If unannealed EBL patterned $Sb_2S_3$ is used instead ($n_{Sb2S3}$=1.790 at 532 nm), the required thicknesses are 0.50 µm in air and 0.87 µm in water, which are only 62% and 35% of PMMA thickness range, which makes the EBL patterned $Sb_2S_3$ competitive replacement of PMMA. Compared with $TiO_2$, the 180 °C annealed EBL patterning $Sb_2S_3$ exhibits very similar refractive index above 633 nm. Thus, amorphous $Sb_2S_3$ can serve as good replacement for $TiO_2$ in metasurface device fabrication. This makes crystalline $Sb_2S_3$ an superior replacement of Si for the optical application in the IR range, especially in the telecommunication C-band (1530-1565 nm). According to the calculation of Wei Jia .ect [19], DOEs composed of $Sb_2S_3$ can provide good performance while still maintaining small thickness. Shuan Qin etal. [20] has reported the simulated performance of metalens composed of $Sb_2S_3$ and Si nanofins based on simulation results.

In this work, we demonstrate the fabrication of 3D FZP and metalens composed of EBL patterned $Sb_2S_3$ for the first time. Fig. 3b and 3c illustrate the schematic of as fabricated 4-level $Sb_2S_3$ FZP and metalens, respectively. The 4-level FZP (Fig. 3b) is functionalized via height induced propagation phase modulation (delay) based on Eq. (2), which is composed of concentric ring staircase structure. The staircase structure shown in Fig. 3f is the translation of the designed 4-level phase profile depicts in Fig. 2a (solid line) based on Eq. (2). The obtained height profile agrees well with the designed height profile (more detailed information is given in Fig. S8a(iv) and S8b). The 4-level metalens (Fig. 3c) is functionalized via

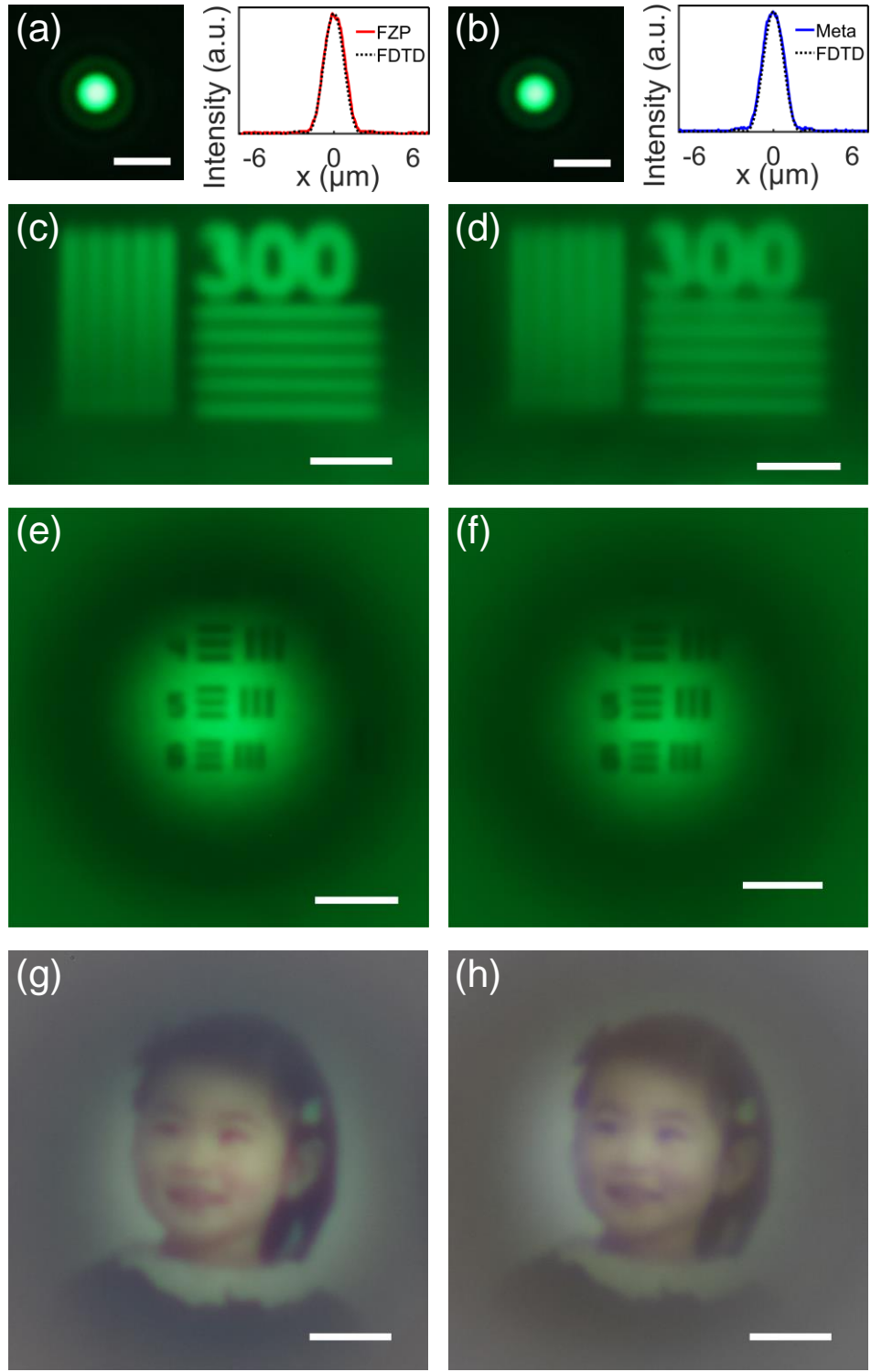

**Figure 4.** Optical performance of 4-level $Sb_2S_3$ (left column) FZP and (right column) metalens characterized with NA = 0.65 objective (40× Olympus PLCN). (a) and (b): Focal spot images (left) and intensity profiles (PSF, right). (c) and (d): Images of TC-RT01 resolution test pattern at 300 μm ($2f_0$) distance away from the lens. The TC-RT01 test pattern was prepared by EBL. The maximum spatial frequency resolved by the $Sb_2S_3$ FZP and metalens is 300 line pairs/mm (with 1.67 μm line distance). (e) and (f): Images of standard

1951 USAF resolution test chart position at 6 cm distance away from the lens. (a-f) measured: at 532 nm vacuum wavelength. (g) and (h): Images taken under white light illumination using a colored positive slide, depicting a portrait image of Wei Wang's daughter. Scale bar: 12 µm for all images.

metaatoms ($Sb_2S_3$ cylindrical pillars) height and diameter (duty cycle) induced propagation phase modulation. Compared with FZP, the design of metalens is much more complicated and computation intensive. The phase delay (give in Fig. S9b) and transmission efficiency (shown in Fig. S9c) maps with height and diameters as sweeping parameters should be calculated (ideally with Rigorous coupled-wave analysis) in the first place. For conventional EBL based metalens fabrication, height modulation of metaatoms is not reported due to fabrication difficulties, thus phase modulation is achieved only in 2D dimension. In this report, the genuine 3D forming capability of g-EBL makes the metalens with height modulation possible. The obtained 3D metalens surface profile (blue line in Fig. 3g) also has good agreement with the designed surface profile (blue line in Fig. 3g). In theory, the whole phase delay map can be utilized for the design of metalens and metasurfaces with g-EBL. As an effort to demonstrate a genuine 3D metalens fabricated with g-EBL, we designed and fabricated a 4-level metalens. Details of the design and fabrication process is given in Supplement Info Part SV. This metalens is designed using the phase delay and transmission maps shown in Fig. S9. According to previous report, metalens with height modulation is only possible with polymer based Two Photon Absorption Lithography (TPAL) [70, 71]. Such 3D metalens offer high focusing efficiency together with achromatic correction in a single OE. However, the spatial resolution of TPAL is only 400 nm [72], which makes it not possible for the metalens fabrication in the VIS range. The refractive index of polymer resist (IP-dip Nanoscibe GmbH) is close to SU-8.

In the end, we demonstrate the optical performance of the aforementioned 4-level $Sb_2S_3$ FZP and metalens (shown in Fig. 3). The schematic illustration of the 4-level phase profile is given in Fig. 3a, with 532 nm working wavelength, a diameter of 57.4 µm, focal length of 150 µm and NA of 0.189 in air. Both OEs were illuminated by a collimated 532 nm center wavelength LED and the focal plane intensity images were magnified using the combination of the objective (NA=0.65) and tube lens (achromatic, $f_0$=500 mm) and captured by a commercial CMOS camera (Canon EOS RP) (see Fig. S10). Fig. 4a and 4b show the measured focal spot (also known as Point Spread Function, PSF), associated intensity profile along x-axis (indicated by the white dashed line) and the Finite Difference Time Domian (FDTD) simulated focal spot intensity. As Fig 4a and 4b show, for both FZP and metalens, the measured PSF match well with the theoretical perdition, thus indicating a negligible wavefront aberration. To demonstrate the use of as fabricated FZP and metalens for practical imaging, we characterized the imaging resolution using the TC-RT01 (Technologie Manufaktur GmbH) and United States Air Force (USAF) 1951 resolution test chart (Thorlabs Inc.) as the target object. The associated measurement configuration is shown in Fig. S11 and S12. The TC-RT01 chart is imaged (Fig.4c and 4d) at $4f_0$ condition, that is to say, with 1:1 ratio magnification. The smallest features of both FZP and metalens are lines with center to center distance of 1.67 µm. The FDTD simulated FWHM is 1.83 µm, which is also in good agreement with experiment result. (USAF 1951 result will be analyzed and given later). Practical usage of $Sb_2S_3$ FZP (Fig. 4g) and metalens (Fig. 4h) is revealed the portrait image of Wei Wang's daughter under white light illumination.

**Part IV: Conclusion** (will be given in the second Arxiv version)

**Part V: Methods** (more details will be given in the second Arxiv version)

Readers can reference to the previous work of WW regarding the EBL patterning procedure [28].

**Part VI: Acknowledgement**

WW thanks for financial support through the IPHT Innovation Project 2021/2022 (690082), DFG, CRC-TRR 234 "CataLight" Project C4. VD appreciates financial support via the DFG Collaborative Research Center SFB 1375 (NOA) Project C02.